\documentclass[doublespacing]{elsart}
\usepackage{graphicx}
\usepackage{ulem}
\usepackage{color}

\journal{Nuclear Instruments and Methods B}
\begin{document}
\begin{frontmatter}
\title{Detecting swift heavy ion irradiation effects with graphene}
\author{O. Ochedowski$^1$},
\author{S. Akc\"oltekin$^1$},
\author{B. Ban--d$`$Etat}$^2$,
\author{H. Lebius}$^2$,
\author{M.~Schleberger$^1$\corauthref{cor}}
\corauth[cor]{Corresponding author:}
\ead{marika.schleberger@uni-due.de}
\address{$^1$Fakult\"at f\"ur Physik and CeNIDE, Universit\"at Duisburg-Essen, 47048 Duisburg, Germany\\
$^2$CIMAP (CEA-CNRS-ENSICAEN-UCBN), 14070 Caen Cedex 5, France}

\begin{abstract}
In this paper we show how single layer graphene can be utilized to study swift heavy ion (SHI) modifications on various substrates. The samples were prepared by mechanical exfoliation of bulk graphite onto SrTiO$_3$, NaCl and Si(111), respectively. SHI irradiations were performed under glancing angles of incidence and the samples were analysed by means of atomic force microscopy in ambient conditions. We show that graphene can be used to check whether the irradiation was successful or not, to determine the nominal ion fluence and to locally mark SHI impacts. In case of samples prepared in situ, graphene is shown to be able to catch material which would otherwise escape from the surface.
\end{abstract}
\end{frontmatter}

\section{Introduction}
Much of the research regarding graphene, a one-atom thick layer of graphite, has been focused on its outstanding electronic properties and its possible role in future semiconducting devices \cite{Bolotin.2008,Kim.2011}, but graphene applications are not restricted to microelectronics. With a mechanical strength of $\sim$100~GPa, a Young modulus of $\sim$1~TPa \cite{Lee.2008} and its impermeability for standard gases as small as He, graphene can be used as an ultra thin protective layer which can e.g.~prevent the oxidation of Cu \cite{Bunch.2008,Chen.2011}. Graphenes huge potential use as a sensor has already been shown for DNA and glucose biosensors \cite{Shao.2010} and even the detection of individual gas molecules has been demonstrated \cite{Schedin.2007}.

Having the ability to reconstruct its perfect hexagonal carbon lattice into pentagons and heptagons near defect sites \cite{Krasheninnikov.2007} and the fact that defects can be exploited to manipulate its physical properties, graphene proofs to be an interesting material for ion irradiation experiments \cite{Pereira.2008}. In previous studies graphene has been irradiated with either low energy ions (LEI) or light ions and with high ion fluences of (from 10$^{11}$~ions/cm$^2$ up to 10$^{18}$~ions/cm$^2$) normal to the surface. These reports have shown that it is indeed possible to induce defects with energetic ions even in suspended graphene sheets \cite{Compagnini.2009,Chen.2009,Tapaszto.2008,Mathew.2011}. Furthermore, it has been shown that by irradiating single layer graphene (SLG) with swift heavy ions (SHI) with $\sim$100~MeV kinetic energy under glancing incidence angle, origami like foldings of the graphene sheet are induced by individual ions \cite{Akcoltekin.2011b}. 

In this paper, we show that graphene can be used to determine whether an irradiation was successful even in the case where the substrate itself is not radiation sensitive. We demonstrate how graphene can then be used to assess either the true incidence angle or the true fluence. Furthermore, the fact that graphene folds upon glancing angle SHI irradiation can be exploited to mark single ion impacts, rendering them accessible to surface sensitive tools. Finally, we show that the mechanical strength of graphene and its impermeability for small molecules make it a perfect catcher for ejected material. 

\section{Experimental}
For our experiments, we prepared graphene on different substrates using the mechanical exfoliation technique of graphite (HOPG - Momentive Performance Materials Quartz, ZYA grade) in order to obtain graphene of the highest quality \cite{Novoselov.2005}. Graphene was exfoliated in ambient conditions onto a freshly cleaved NaCl(100) single crystal (Korth Kristalle), referred to as G/NaCl in the following, and SrTiO$_3$(100) (Crystec), referred to as G/STO. Additionally, single layer graphene was prepared under ultra high vacuum (UHV) conditions on a Si(111)7x7 substrate (G/Si). For the latter, a Si(111) wafer (n-doped, 10-20 Ohm$\cdot$cm) was repeatedly flash heated to 1250$^{\mathrm{o}}$C and the quality of the reconstruction was confirmed by low energy electron diffraction (LEED). For more information regarding this preparation procedure see \cite{Ochedowski.2012}.

SHI irradiation took place at the IRRSUD beamline (GANIL, Caen - France). The incidence angles ranged from $\Theta=$1$^{\mathrm{o}}$ to 3$^{\mathrm{o}}$ with respect to the sample surface and the ion fluence on the samples were adjusted to around (5--10)~ions/cm$^2$ to be able to analyze single, non-overlapping SHI impact events. For the irradiations Xe$^{23+}$ and Pb$^{28+}$ ions were used. The respective stopping power in the substrate materials was in the range of $S_{e}=$(12--19)~keV/nm, in an amorphous carbon target with the density of graphite it would be $S_{e}=$(15--17)~keV/nm. These values were calculated using the SRIM 2008 software package \cite{Ziegler.2010}.

%Different SHI have been used for the experiments, them being Xe$^{23+}$, Ta$^{24+}$, Pb$^{28+}$ and Pb$^{31+}$. The respective stopping power in the substrate materials was in the range of $S_{e}=$(12--20)~keV/nm, in an amorphous carbon target with the density of graphite it was $S_{e}=$()~keV/nm. These values were calculated using the SRIM software package \cite{Ziegler.2010}.

After the irradiation experiments, the samples were analysed in ambient with an atomic force microscope (AFM, Veeco Dimension 3100 system).
Measurements were performed in tapping mode using Nanosensors NCHR cantilever with typical first resonant frequencies of $f_{0}\sim$300~kHz.
 
\section{Results and Discussion}
\begin{figure}[b]
	\centering
	\includegraphics[width=\columnwidth]{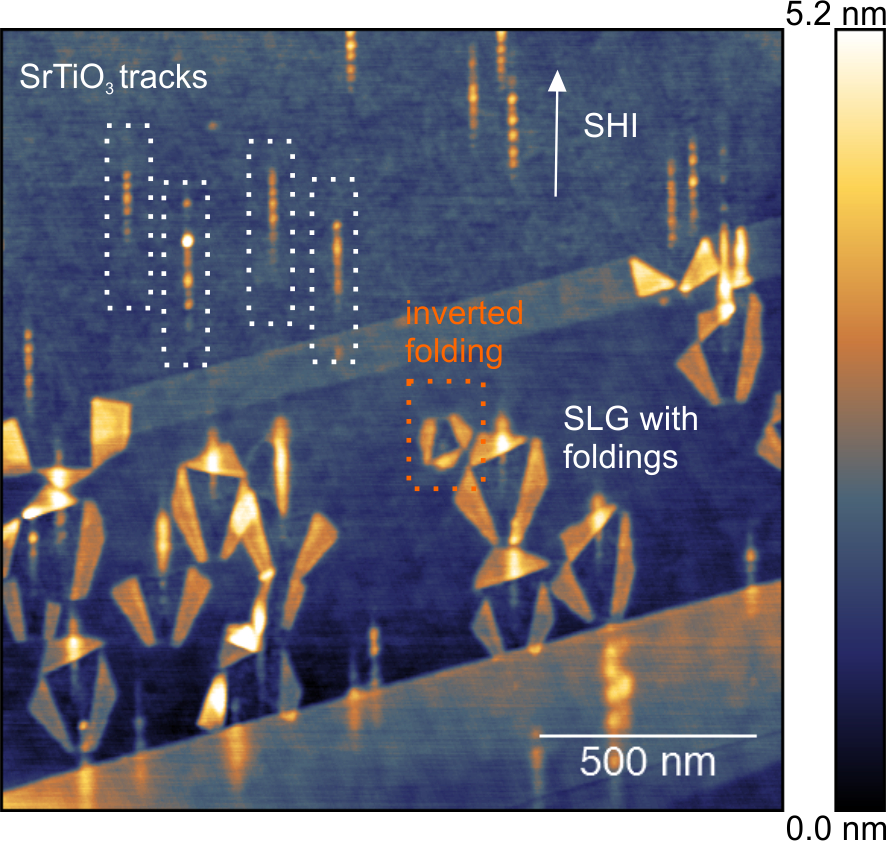}
	\caption{AFM topography of G/STO showing the typical graphene foldings induced by SHI irradiation (91 MeV $^{129}$ Xe$^{23+}$; $S_{e}$=19.0~keV/nm ; $\Theta=2.6^{\mathrm{o}}$) under glancing incidence. Several surface tracks on the SrTiO$_{3}$ substrate are marked by white boxes. Folded areas on SLG are connected to surface tracks in SrTiO$_{3}$ underneath. In the orange box, a rare inverted folded structure can be observed.}
	\label{Figure1}
\end{figure}

We begin with the typical image observed when irradiating an exfoliated graphene flake on an insulator, see fig.~\ref{Figure1}. Here, G/STO was irradiated by 91 MeV $^{131}$ Xe$^{23+}$ SHI with a stopping power of $S_{e}$=19.0~keV/nm under a glancing incidence angle of 2.6$^{\mathrm{o}}$. Upon impact of the SHI, graphene is folded in a typical pattern. The exact shape of the pattern may vary depending on different parameters like incidence angle and target material \cite{Akcoltekin.2011}. At this angle a typical folding consists of two to three folded areas, two of them alongside and symmetrical with respect to the ion beam trajectory, and one of them oriented almost perpedicular to the track (and usually found downstream, see fig.~\ref{Figure1}). The mechanism of the folding has been explained in detail elsewhere \cite{Akcoltekin.2011b}, here we will give only a brief description.

SHI irradiation of insulators like SiO$_2$ or SrTiO$_{3}$ under glancing incidence is known to create surface tracks of molten material along the ion trajectory \cite{Akcoltekin.2007,Akcoltekin.2009}. As this state is rapidly quenched, a chain of nanosized hillocks remains on the surface. If the track area is covered by graphene, this hot surface material pushes through the graphene which has been damaged along the track beforehand to some extent by the projectile itself. The SrTiO$_3$ surface track can be seen in the exposed area between the folded graphene (see fig.~\ref{Figure1}). After the projectile has reached a certain depth, the surface track in SrTiO$_{3}$ continues underneath an otherwise intact graphene sheet. Note, that virtually every surface track in G/STO observed on single layer graphene has a folding attached to it. Thus, the sensitivity of SLG to SHI induced damage is 100\%. Folding patterns deviating from the regular one, like inverted foldings (see orange box in fig.~\ref{Figure1}), can be sometimes observed, too. These foldings are most likely caused by elastically scattered SHI, where the trajectory is not a straight line.

\begin{figure}
	\centering
	\includegraphics[width=\columnwidth]{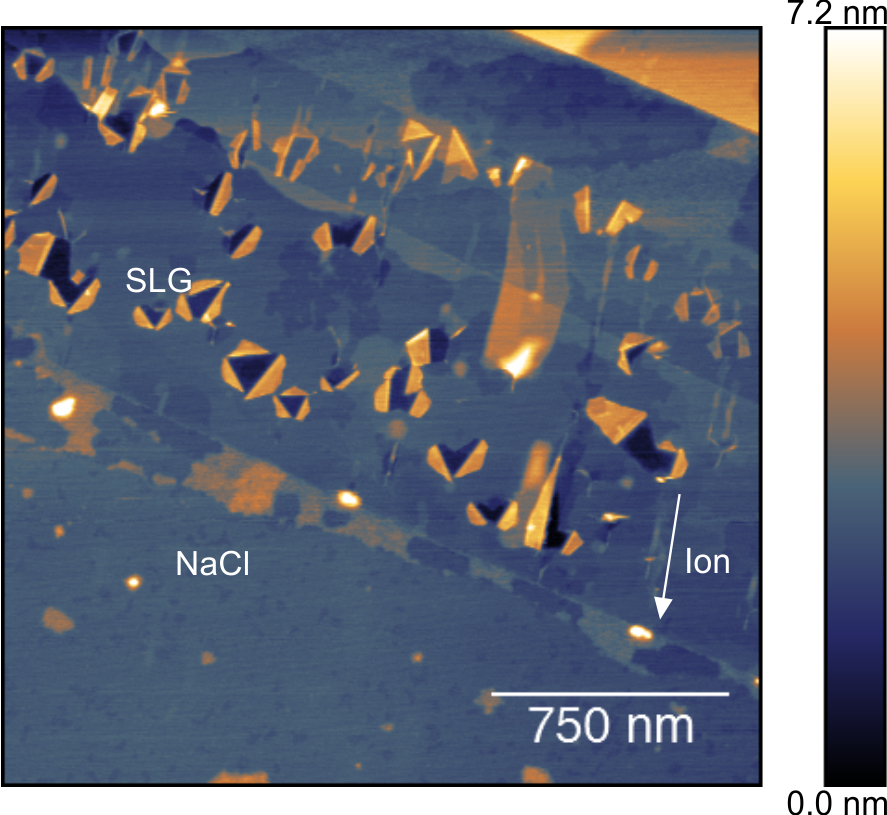}
	\caption{AFM topography of SLG on NaCl, irradiated with SHI (104 MeV $^{207}$Pb$^{28+}$; $S_{e}$=12.77 keV/nm; $\Theta=2.86^{\mathrm{o}}$). SLG shows typical folded areas on the surface while NaCl shows no signs of modification.}
	\label{Figure2}
\end{figure}

Until now, SHI irradiation induced foldings in graphene have been studied exclusively on materials which show distinct surface tracks \cite{Akcoltekin.2011b,Akcoltekin.2011}. There are however materials, on which no surface tracks or morphological modifications are expected \cite{Itoh.2009} or detected, like e.g.~Si \cite{SAkcoltekin.2008}. Here we demonstrate, that on substrates which show no surface tracks in form of protrusions or nano-sized hillocks, graphene foldings can still occur, as can be seen in fig.~\ref{Figure2}. Fig.~\ref{Figure2}(a) shows an image taken from G/NaCl which was irradiated with 104~MeV $^{207}$Pb$^{28+}$ under an incidence angle of $\Theta=3^{\mathrm{o}}$. The NaCl surface at this resolution shows no signs of modification due to the SHI impact while the SLG flake shows the typical SHI induced foldings. Thus, graphene can here be used to detect SHI impacts, localize and count them (fluence control). 

We determine the zero degree grazing angle by optically aligning the surface parallel to the beam direction by means of a telescope aligned on the optical axis of the beam line. A stepping motor with a repetability of about 5/100 of a degree is used to turn the target to the nominal angle. In light of the extreme good repeatability of the rotation, the main error bar stems from the optical alignment, and we expect an error bar of about 0.2~degree. If the efficiency of damage is known as in the case of G/STO, the true angle of incidence can be calibrated by comparing the experimentally determined fluence with the nominal one: $\Theta_e=arcsin(\mathrm{experimental~fluence}/\mathrm{nominal~fluence})$. For G/NaCl, the efficiency is unknown, but if we assume 100\% for the sample shown in fig.~\ref{Figure2} we determine $\Theta_e=2.86^{\mathrm{o}}$ which is very close to the nominal one of $\Theta_n=3^{\mathrm{o}}$. 

In addition, graphene offers an efficient way to search for possible irradiation damage of the NaCl surface. One could now simply limit the search to the areas marked by foldings and study those areas inside of the graphene and/or foldings located at the edge of the flake in more detail. This allows for smaller scanning windows with the possibility of enhanced resolution.
 
\begin{figure}
	\centering
	\includegraphics[width=\columnwidth]{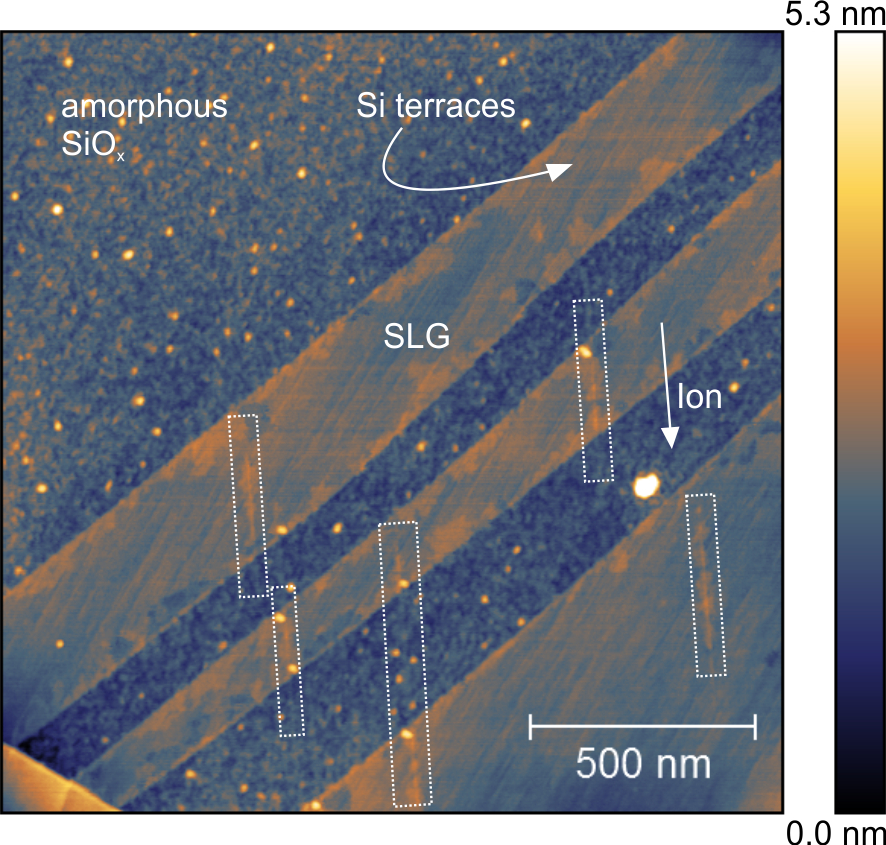}
	\caption{AFM topography of SLG on Si(111), irradiated with 92 MeV $^{129}$Xe$^{23+}$; $S_{e}$=12.35 keV/nm; $\Theta=1^{\mathrm{o}}$. The sample was prepared by exfoliating graphene in situ onto a Si(111)7x7 substrate. After exposure to ambient bare Si is directly oxidized whereas graphene covered Si still shows terrace steps. Surface tracks on this sample are observed exclusively on the graphene covered Si substrate, marked by white boxes.}
	\label{Figure3}
\end{figure}

As the mechanical exfoliation technique of graphene is generally used under ambient atmosphere, all kind of adsorbates like oxygen, nitrogen or water will be enclosed between graphene and the substrate forming an undefined interface layer (IFL). To study the yet unkown influence of this IFL on the SHI irradiation, a graphene flake was exfoliated onto a Si(111)7x7 substrate in situ. After the in situ preparation, the sample was exposed to ambient and analysed with AFM and Raman spectroscopy. This showed, that while a $\sim$1.5~nm native SiO$_{x}$ layer is formed on top of Si, graphene covered Si still exhibits terrace steps (see \cite{Ochedowski.2012} and fig.\ref{Figure3}). The latter indicates that oxidation, and thus the formation of an IFL, might have been successfully prevented. An AFM topography image of this sample after irradiation with 92 MeV $^{129}$Xe$^{23+}$ SHI is shown in fig.\ref{Figure3}.

On the amorphous SiO$_{x}$ surface no signs of ion induced surface tracks can be observed. The native oxide is obviously too thin to give rise to the faint surface tracks typical for bulk SiO$_2$ samples \cite{Akcoltekin.2009}. Nevertheless, the ion solid interaction can still be detected because on the graphene covered Si, distinct surface tracks with an average height of about 0.4~nm are measured. Hence, graphene does not fold with a largely absent IFL despite surface tracks being formed underneath. This finding shows that in addition to the surface track from the substrate, the IFL must play a role for the folding mechanism.

The exact nature of the tracks detected in G/SiO remains unclear. However, from our data it is obvious that graphene retains material which otherwise would have left the surface. The use of graphene as a protective layer for ion irradiation experiments with gas targets has already been proposed by Lehtinen et al. \cite{Lehtinen.2010}. Stolyarova et al.~observed the formation of graphene bubbles after irradiating graphene on SiO$_{2}$ with 0.4-0.7 MeV protons normal to the surface \cite{Stolyarova.2009}. From fig.~\ref{Figure3} it can be clearly seen that the use of graphene as a protective layer does not stop at retaining gaseous materials but can be extended to catching ejected solid material. 

\section{Conclusion}
In this work we have shown, that ultrathin sheets of graphene can be used to detect and study SHI irradiation effects in various ways. Foldings can be used as impact markers enabling calibration of irradiation parameters as well as targeting irradiated areas in materials without distinct track formation. The ability of graphene to catch ejected material may open a new way to implement foreign material into graphene as well as to study ion induced modifications which are otherwise transient.

\section*{Acknowledgement}
This work has been supported by the European Community as an Integrating Activity Support of Public and Industrial Research Using Ion Beam Technology (SPIRIT) under EC contract no. 227012 and by the German Science Foundation (SPP 1459: Graphene and SFB 616: Energy dissipation at surfaces). The experiments were performed at the IRRSUD beamline of the Grand Accelerateur National d$`$Ions Lourds GANIL, Caen, France.

\newpage
\section*{Figures}

\end{document}